\documentclass[superscriptaddress,preprint,amsmath,amssymb,aps,prd,longbibliography]{revtex4-1}

\usepackage{xcolor}
\usepackage[utf8]{inputenc}
\usepackage{array}
\usepackage{makecell}
\usepackage{tabularx}
\usepackage{pifont}
\usepackage{graphicx}
\usepackage{appendix}

\newcommand{\cmark}{\ding{51}}
\newcommand{\xmark}{\ding{55}}

\begin{document}

\title{Stochastic Schr\"{o}dinger equations and conditional states: a general Non-Markovian quantum electron transport simulator for THz electronics}


\author{Devashish Pandey}
\affiliation{Departament d'Enginyeria Electr\`onica. Universitat Aut\`onoma de Barcelona, 08193-Bellaterra (Barcelona), Spain}
\author{Enrique Colom\'es}
\affiliation{Departament d'Enginyeria Electr\`onica. Universitat Aut\`onoma de Barcelona, 08193-Bellaterra (Barcelona), Spain}
\author{Guillermo Albareda}
\email{guillermo.albareda@mpsd.mpg.de}
\affiliation{Departament d'Enginyeria Electr\`onica. Universitat Aut\`onoma de Barcelona, 08193-Bellaterra (Barcelona), Spain}
\affiliation{Max Planck Institute for the Structure and Dynamics of Matter, 22761 Hamburg, Germany}
\affiliation{Institute of Theoretical and Computational Chemistry, Universitat de Barcelona. 08028 Barcelona. Spain}
\author{Xavier Oriols}
\email{xavier.oriols@uab.cat}
\affiliation{Departament d'Enginyeria Electr\`onica. Universitat Aut\`onoma de Barcelona, 08193-Bellaterra (Barcelona), Spain}



\begin{abstract}
A prominent tool to study the dynamics of open quantum systems is the reduced density matrix. 
Yet, approaching open quantum systems by means of state vectors has well known computational advantages. 
In this respect, the physical meaning of the so-called conditional states in Markovian and non-Markovian scenarios has been a topic of recent debate in the construction of stochastic Schr\"{o}dinger equations.
We shed light on this discussion by acknowledging the Bohmian conditional wavefunction as the proper mathematical object to represent, in terms of state vectors, an arbitrary subset of degrees of freedom. 
As an example of the practical utility of these states, we present a time-dependent quantum Monte Carlo algorithm to describe electron transport in open quantum systems under general (Markovian or non-Markovian) conditions. By making the most of trajectory-based and wavefunction methods, the resulting simulation technique extends, to the quantum regime, the computational capabilities that the Monte Carlo solution of the Boltzmann transport equation offers for semi-classical electron devices.
\end{abstract}


\maketitle

\tableofcontents

\section{Introduction}
\label{intro}
Thanks to its accuracy and versatility, the Monte Carlo solution of the Boltzmann transport equation has been, for decades, the preferred computational tool not only to predict the DC but also AC, transient and noise performances of semi-classical electron devices~\cite{jacoboni1983monte}.  
In the past decade, however, due to the miniaturization of electronic devices (with active regions approaching the de Broglie wavelength of the transport electrons), a majority of the device modeling community has migrated from semi-classical to fully quantum simulation tools, marking the onset of a revolution in the community devoted to semiconductor device simulation. Today, a number of quantum electron transport simulators are available to the scientific community~\cite{nemo,nextnano,tibercad,nanocad,transiesta}. The amount of information that these simulators can provide, however, is mainly restricted to the stationary regime and therefore their predicting capabilities are still far from those of the traditional Monte Carlo solution of the semi-classical Boltzmann transport equation~\cite{jacoboni1983monte}. This limitation poses a serious problem in the near future, as electron devices are foreseen to operate at the Terahertz (THz) regime. At these frequencies, the discrete nature of electrons in the active region is expected to generate unavoidable fluctuations of the current that could interfere with the correct operation of such devices both for analog and digital applications~\cite{Albareda_2009}. 

A formally correct approach to electron transport beyond the quasi-stationary regime lies on the description of the active region of an electron device as an open quantum system~\cite{breuer2002theory,smirne2010initial}.
As such, one can then borrow any state-of-the-art mathematical tool developed to study open quantum systems~\cite{de2017dynamics,PhysRevA.78.022112}. A preferred technique has been the stochastic Schr\"{o}dinger equation (SSE) approach~\cite{gisin1989stochastic,pearle1989combining, carmichael2009open,van1992stochastic,de2011non,goetsch1994linear,gatarek1991continuous,gambetta2002non}. Instead of directly solving equations of motion for the reduced density matrix, the SSE approach exploits the state vector nature of the so-called conditional states to alleviate some computational burden (and ensuring complete positive map by construction~\cite{rivas2014quantum}). Even if this technique allows to reconstruct the full density matrix in any circumstance, a discussion on whether dynamical information can be directly extracted from such conditional states in non-Markovian scenarios has appeared recently in the literature~\cite{diosi2008non,wiseman2008pure}. This debate is very relevant to us as we are interested in computing not only one-time expectation values (i.e. DC performance), but also dynamical properties (i.e. AC, transient and noise), such as multi-time-correlation functions, at THz frequencies. At this frequencies the environment correlations are expected to decay on a time-scale comparable to the time-scale relevant for system evolution~\cite{eisenberg2017dynamics}. Furthermore, the displacement current becomes important at very high frequencies and a self-consistent solution of the Maxwell equations and the time-dependent Schr\"{o}dinger equation is necessary~\cite{eisenberg2017dynamics,oriols2013quantum}.

Some light on how to utilize the SSE technique to access dynamical information without the need of reconstructing the reduced density matrix has been already shed by Wiseman and Gambetta by acknowledging the Bohmian conditional wavefunction as the proper mathematical tool to describe general open quantum systems in non-Markovian scenarios~\cite{gambetta2003interpretation,gambetta2003modal}. In this work we reinforce this idea by showing that the Bohmian conditional wavefunction is an exact decomposition and recasting of the unitary time-evolution of a closed quantum system that yields a set of coupled, non-Hermitian, equations of motion that allows to describe the evolution of arbitrary subsets of the degrees of freedom on a formally exact level. Furthermore, since the measurement process is defined as a routine interaction between subsystems in Bohmian mechanics, conditional states can be used to describe either the measured or unmeasured dynamics of an open quantum system.
As an example of the practical utility of the conditional wavefunctions, we present a Monte Carlo simulation scheme to describe quantum electron transport in open systems that is valid both for Markovian or non-Markovian regimes and that guarantees a dynamical map that preserves complete positivity~\cite{colomes2017quantum,PhysRevLett.98.066803,alarcon2013computation,albareda2010time,albareda2009many}.

This paper is structured as follows. In section~\ref{electron_devices} we provide a brief account on the SSE approach and on how nanoscale electron devices can be understood as open quantum systems. Section~\ref{condionalstates_interpretation} focuses on the physical interpretation of the conditional states (i.e., system states conditioned on a particular value of the environment) in the contexts of the Orthodox and Bohmian quantum mechanical theories. Section~\ref{BITLLES} provides an overall perspective on the points raised in the previous sections and puts into practice the conditional wavefunction concept to build a general purpose electron transport simulator, called BITLLES, beyond the steady state (Markovian) regime. As an example of the use of conditional states, numerical simulations of the THz current in a graphene electron device are presented in Section~\ref{nm}. Final comments and conclusions are indicated in Section~\ref{conclusion}.

\section{Electron devices as open quantum systems}
\label{electron_devices}

In this section we introduce the SSE approach to open quantum systems and discuss how it can be used to reconstruct the reduced density matrix. We then explain how a nanoscale electron device can be understood as an open quantum system and how the SSE approach can be applied to predict its performance. 

\subsection{Open quantum systems}

\begin{figure}
\centering
\includegraphics[width=0.7\textwidth]{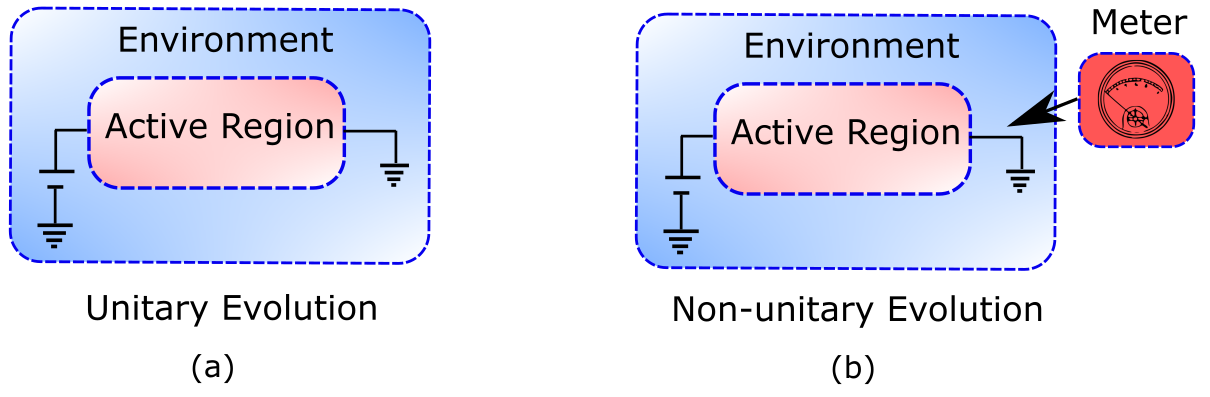}
\caption{Panel (\textbf{a}): Schematic representation of an open quantum system, which can be partitioned into active region and environment. The evolution of the entire device is described by the state $|\Psi(t)\rangle$ that evolves unitarily according to the time-dependent Schr\"odinger equation. Panel (\textbf{b}): Schematic representation of a measured open quantum system, which can be partitioned into meter, active region and environment. The evolution of the device plus environment wavefunction is no longer unitary due to the (backaction) effect of the measuring apparatus.}
\label{schr}
\end{figure}

As it is usual we start with a closed quantum system (as the one shown in Figure~\ref{schr}a). This system is represented by a pure state, $|\Psi(t)\rangle$, which evolves unitarily according to the time-dependent Schr\"odinger equation
\begin{equation}\label{Scho}
    i\hbar \frac{\partial |\Psi(t)\rangle}{\partial t} = \hat H |\Psi(t)\rangle.
\end{equation}
Finding a solution to Equation~\eqref{Scho} is inaccessible for most practical scenarios due to the large number of degrees of freedom involved. Therefore, it is a common practice to partition the system into two subsets of degrees of freedom, viz., open system and environment \cite{breuer2002theory}. The open system can be then described by a reduced density matrix 
\begin{equation}
  \hat\rho_\text{sys}(t)= \text{Tr}_\text{env}\left[|\Psi(t)\rangle\langle\Psi(t)|\right ],
  \label{reduced}
\end{equation}
where $\text{Tr}_\text{env}$ denotes the trace over the environment degrees of freedom.
The reduced density matrix $\hat\rho_\text{sys}$ can be shown to obey, in most general circumstances, a non-Markovian master equation~\cite{nakajima1958quantum, zwanzig1960ensemble}:
\begin{equation}
\frac{\partial \hat\rho_\text{sys}(t)}{\partial t}=-i\left[\hat H_\text{int}(t), \hat\rho_\text{sys}(t)\right] + \int_{t_0}^t \hat{\mathcal{K}}(t,t') \rho_\text{sys}(t') dt'
\label{nonmarkov}
\end{equation}
where $\hat H_\text{int} (t)$ is a system Hamiltonian operator in some interaction picture and $\hat{\mathcal{K}}(t,s)$ is the ``memory time'' superoperator, which operates on the reduced state $\hat\rho_\text{sys}(t)$ and represents how the environment affects the system. If the solution to Equation~\eqref{nonmarkov} is known then the expectation value of any observable $\hat A$ of the system can be evaluated as:
\begin{equation}
 \langle \hat A(t) \rangle = \text{Tr}_\text{sys}[\hat{\rho}_\text{sys}(t) \hat A],
 \label{measure}
\end{equation}

Unfortunately, solving Equation~\eqref{nonmarkov} is not an easy task. The effect of $\hat{\mathcal{K}}(t,s)$ on $\hat\rho_\text{sys}(t)$ cannot be explicitly evaluated in general circumstances. Moreover, even if the explicit form of $\hat{\mathcal{K}}(t,s)$ is known, the solution to Equation~\eqref{nonmarkov} is very demanding as the density matrix $\hat\rho_\text{sys}(t)$ scales very poorly with the number of degrees of freedom of the open system. Finally, if one is aiming at computing multi-time correlations functions, then it is necessary to incorporate the effect (backaction) of the successive measurements on the evolution of the reduced density matrix, which is, in general non-Markovian regimes, a very complicated task, both from the practical and conceptual point of views.

\subsection{Stochastic Schrodinger equations}

A breakthrough in the computation of the reduced density matrix in Equation~\eqref{reduced} came from the advent of the SSE approach~\cite{dalibard1992wave}. The main advantage behind the SSE approach is that the unknown to be evaluated is in the form of a state vector (of dimension $N_\text{sys}$) rather than a matrix (of dimension $N_\text{sys}^2$) and thus there is an important reduction of the associated computational cost. In addition, it provides equations of motion that, by construction, ensure a complete positive map~\cite{rivas2014quantum} so that the SSE approach guarantees that the density matrix always yields a positive probability density, a requirement that is not generally satisfied by other approaches that are based on directly solving Equation~\eqref{nonmarkov} \cite{strunz1999open}.  

The central mathematical object in the SSE approach to open quantum systems is the conditional state of the system:
\begin{equation}
 |\psi_{q}(t) \rangle =\frac{\left(\langle q| \otimes \hat I_\text{sys}\right) |\Psi(t)\rangle}{\sqrt{P(q,t)}},
 \label{cond}
\end{equation}
where $P(q,t)= \langle \psi_{q}(t) |\psi_{q}(t) \rangle =\langle \Psi(t)|\hat I_\text{sys}\otimes | q\rangle\langle q| \otimes \hat I_\text{sys} |\Psi(t)\rangle$ and $|q\rangle$ are the eigenstates  of the so-called unraveling observable $\hat Q$ belonging to the Hilbert space of the environment. To simplify the discussion, and unless indicated,  $q$  represents the collection of degrees of freedom of the environment in  a single variable. Using the eigenstates $|q\rangle$ as a basis for the environment degrees of freedom, it is then easy to rewrite the full state $|\Psi(t)\rangle$ as:
\begin{equation}
  |\Psi(t)\rangle = \int dq \sqrt{P(q,t)} |q\rangle \otimes |\psi_{q}(t) \rangle,
  \label{decom}
\end{equation}
which can be simply understood as a Schmidt decomposition of a bipartite (open system plus environment) state. Thus, a complete set of conditional states can be always used to reproduce the reduced density matrix at any time as:
\begin{eqnarray}
    \hat{\rho}_\text{sys}(t) =  \int dq  P(q,t) |\psi_{q}(t)\rangle \langle\psi_{q}(t)|.
    \label{result}
\end{eqnarray}
Let us note that no specific (Markovian or non-Markovian) assumption was required to write Equation~\eqref{result}. In fact, the above definition of the reduced density matrix simply responds to the global unitary evolution in Equation~\eqref{Scho}, which (as depicted in Figure \ref{schr}a) does not include the effect of any measuring apparatus.  

\subsection{Nanoscale electron devices as open quantum systems}

At first sight, one could be inclined to say that a nanoscale electron device perfectly fits into the above definition of open quantum system. Then, the open system would be the device's active region and the environment (including the contacts, the cables, ammeter etc.) would be called reservoirs or contacts (see Figure~\ref{schr}a). In addition, the observable of interest $\hat A$ in Equation~\eqref{measure} would be, most probably, the current operator $\hat I$. 
As long as we are interested only on single-time expectation values, i.e., static or stationary properties, this picture (and the picture in Figure~\ref{schr}a) is perfectly valid. Therefore, the SSE approach introduced in Equations~\eqref{cond}, \eqref{decom} and \eqref{result} can be easily adopted to simulate electron devices and hence predict their performance. 

However, if one aims at computing dynamical properties such as time-correlation functions, e.g., $\langle  I(t+\tau)I(t) \rangle$, then a valid question is whether such an expectation value is expected to be measurable at the laboratory. And if so, what would be then the effect of the measurement of $I$ at time $t$ on the measurement of $I$ at a later time $t+\tau$?. Figure~\ref{schr}b schematically depicts this question by drawing explicitly the measuring apparatus (or meter). As it is well known, the action of measuring in quantum mechanics is not innocuous. Quite the opposite, in many relevant situations, extracting information from a system at time $t$ has a non-negligible effect on the subsequent evolution of the system and hence also on what is measured at a later time $t+\tau$. 
Therefore, as soon as we are concerned about dynamics information (i.e., time-correlation functions), we need to ask ourselves whether an approach to open quantum systems such as the SSE approach can be of any help.
In the next section we will make an effort to answer this question and to understand whether the conditional states $|\psi_{q}(t)\rangle$ defined in Equation~\eqref{cond} do take into account the backaction of the measuring apparatus depicted in Figure~\ref{schr}b.


\section{Interpretation of conditional states in open quantum systems}
\label{condionalstates_interpretation}

The conditional states in Equation~\eqref{cond} were first interpreted as a simple numerical tool~\cite{dalibard1992wave}, that is, exploiting the result in Equation~\eqref{result} as a numerical recipe to evaluate any expectation value of interest. This interpretation is linked to the assumption that the operator $\hat A$ in Equation~\eqref{measure} is the physically relevant operator (associated to a real measuring apparatus), while the operator $\hat Q$ associated to the definition of the conditional state in Equation~\eqref{cond} is only a mathematical object with no attached physical reality. 
In more recent times, however, it has been generally accepted that the conditional states in Equation~\eqref{cond} can be interpreted as the states of the system conditioned on a type of sequential (sometimes referred as continuous) measurement~\cite{barchielli2012quantum} of the operator $\hat Q$ of the environment (as a physical measuring apparatus that substitutes the no longer needed operator $\hat A$)~\cite{carmichael2009open,wiseman2009quantum,breuer2002theory}. From  a practical point of view, this last interpretation is very attractive as it would allow to link the conditional states, $|\psi_{q}(t) \rangle$, at different times and thus compute time-correlation functions without the need of introducing the measuring apparatus. Whether or not this later interpretation is physically sound in general circumstances is the focus of our discussion in the next subsections.

\subsection{The Orthodox interpretation of conditional states}

\begin{figure}
\centering
\includegraphics[width=\textwidth]{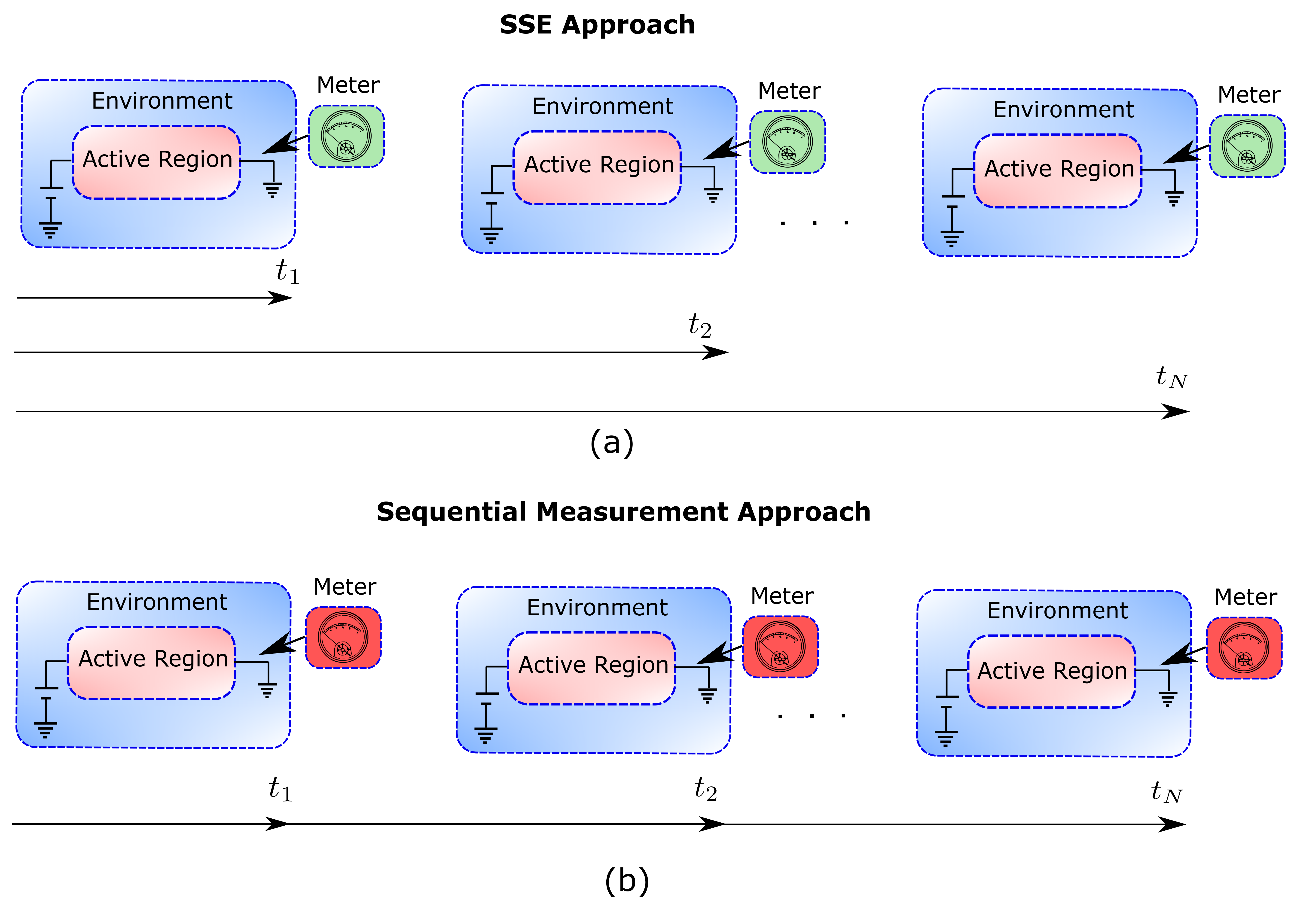}
\caption{Panel (\textbf{a}): Schematic representation of the SSE approach. The states of the system conditioned on a particular value of the environment at time $t$, $|\psi_{q}(t) \rangle$, can be given a physical meaning only if no measurement has been performed at a previous time $t'<t$. This approach can be always used to reconstruct the correct reduced density matrix of the system at any time, but cannot be used to link in time the conditional states for non-Markovian scenarios.
Panel (\textbf{b}): Schematic representation of a sequential measurement. The wavefunction of the system plus environment is measured sequentially. In this picture, the link between the states of the full system plus environment at different times is physically motivated.}
\label{SSE_fig}
\end{figure}

Let us start by discussing, in the orthodox quantum mechanics theory, what is the physical meaning of the conditional states that appear in Equation~\eqref{cond}. When the full closed system follows the unitary evolution of Figure~\ref{schr}a, then the conditional state $|\psi_{q}(t)\rangle$ can be understood as the (renormalized) state that the system is left after projectively measuring the property $Q$ of the environment (with outcome $q$).
This can be easily seen by noting that the superposition in Equation~\eqref{decom} is, after a projective measurement of $Q$, reduced (or collapsed) to the product state
\begin{equation}
    |{\Psi}_{q}(t)\rangle = \sqrt{P(q,t)} |q\rangle \otimes |\psi_{q}(t)\rangle.
\end{equation} 
It is important to notice that the conditional state $|\psi_{q'}(t') \rangle$ at a later time, $t'>t$, can be equivalently defined as the state of the system when the superposition in Equation~\eqref{decom} is measured at time $t'$ and yields the outcome $q'$. This interpretation, however, is only valid if no previous measurement (in particular at $t$) has been performed, as depicted in Figure \ref{SSE_fig}a. Otherwise, the collapse of the wavefunction at time $t$, yielding the state $\sqrt{P(q,t)} |q\rangle \otimes |\psi_{q}(t)\rangle$, should be taken into account in the future evolution of the system, which would not be the same as if the measurement had not been performed at the previous time. 
Therefore, the equation of motion of the conditional states, as defined in Equation~\eqref{cond}, cannot be, in general, the result of a sequential measurement protocol such as the one depicted in Figures~\ref{schr}b or \ref{SSE_fig}b. This conclusion seems obvious if one recalls that our starting point was Figure~\ref{schr}a, where there is no measurement.

\subsubsection{Orthodox conditional states in Markovian scenarios}
\label{OCS}

Even if the conditional states solution of the SSE cannot be generally interpreted as the result of a sequential measurement, such an interpretation has been proven to be very useful in practice for scenarios that fulfill some specific type of Markovian conditions.  
We are aware that there is still some controversy on how to properly define Markovianity in the quantum regime (see, e.g., Ref.~\cite{rivas2014quantum}), so it is our goal here only to acknowledge the existence of some regimes (i.e., particular observation time intervals) of interest where the role of the measurement of the environment has no observable effects. In this regime,  Figures~\ref{schr}a and ~\ref{schr}b as well as Figures~\ref{SSE_fig}a and ~\ref{SSE_fig}b can be thought as being equivalent. 

\begin{figure}
\centering
\includegraphics[width=0.5\textwidth]{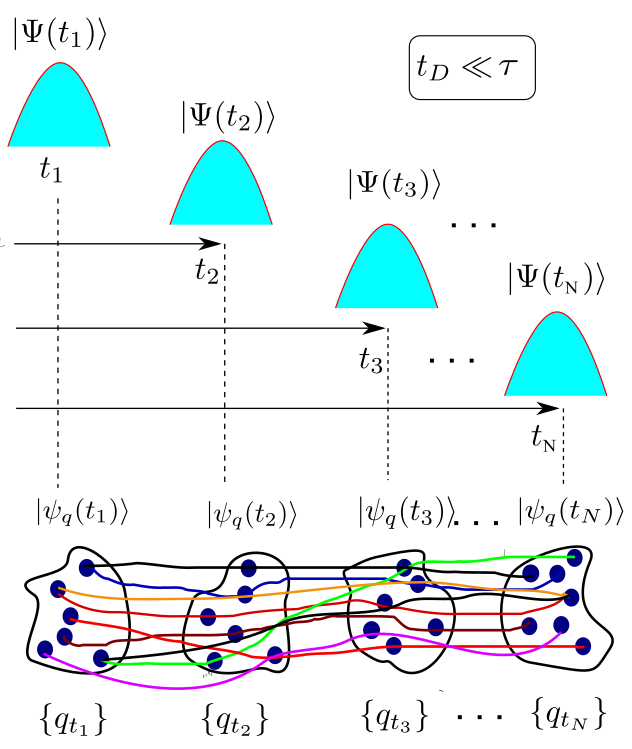}
\caption{Schematic representation of the combined system plus environment wavefunction (blue gaussians) measured at different times that result in a state of the system $|\psi_{q}(t)\rangle$ conditioned to the set of environment values $\{q_t\}$ shown in dark blue circles. In the Markovian regime there exists no specific recipe about how the different $q$'s must be connected in time (colored solid lines). No matter how these points are connected in time, one always gets the right expectation value in Equation~\eqref{corr_funct_markov}.}
\label{mark}
\end{figure}

In our pragmatical definition of Markovianity, the entanglement between system and environment decays in a time scale $t_D$ that is much smaller than the observation time interval $\tau$, i.e., $t_D \ll \tau$.
In this regime, the environment itself can be thought of as a type of measuring operator (as appears in generalized quantum measurement theory~\cite{kraus1983states}) that keeps the open system in a pure state after the measurement. The open system can be then seen as an SSE in which the stochastic variable $q_t$ (sampled from the distribution $P(q_t,t)$) is directly the output of a sequential measurement of the environment. The stochastic trajectory of this conditioned system state generated by the
(Markovian) SSE is often referred to as a quantum trajectory~\cite{carmichael2009open,wiseman2009quantum,breuer2002theory} and can be used, for example, to evaluate 
time-correlation functions of the environment as:
\begin{equation}\label{corr_funct_markov}
  \langle Q(t) Q(t+\tau) \rangle \stackrel{t_D \ll \tau}{=} 
  \int\int P(q_t,t) P(q_{t+\tau},t+\tau) q_t q_{t+\tau} dq_t dq_{t+\tau}=\langle Q(t)\rangle \langle Q(t+\tau) \rangle .
\end{equation}
Let us emphasize that the stochastic variables $q_t$ and $q_{t+\tau}$ in Equation~\eqref{corr_funct_markov} are sampled, separately, from the probability distributions $P(q_t,t) = \langle \psi_{q}(t) |\psi_{q}(t)\rangle$ and $P(q_{t+\tau},t+\tau) = \langle \psi_{q}(t+\tau) |\psi_{q}(t+\tau)\rangle$. Therefore, as we have schematically depicted in Figure~\ref{mark}, no matter how the trajectories $\{q_t\}$ are connected in time, one always obtains the correct time-correlation function $\langle Q(t) Q(t+\tau) \rangle$.

It is important to realize that we started our discussion on the physical meaning of the Markovian SSE with an open system whose environment is not being measured (see Figures~\ref{schr}a and ~\ref{SSE_fig}a). Noticeably, we have ended up discussing about an environment that is being measured every time interval $\tau$ (see Figure~\ref{SSE_fig}b). How is that possible? Well, the reason is that measuring the environment at time $t$ does not affect the system conditional states at a later time $\tau$ when the built in correlations in the environment due to the measurement at time $t$ decay in a time interval $t_D$ much smaller than the time between measurements $\tau$. Therefore, Figures~\ref{schr}a and ~\ref{schr}b as well as Figures~\ref{SSE_fig}a and \ref{SSE_fig}b are not distinguishable when $t_D\ll\tau$. In this sense, the Markovian regime has some similarities with a classical system, where it is accepted that information can be extracted without perturbation. 

\subsubsection{Orthodox conditional states in non-Markovian scenarios}

For nanoscale devices operating at very high (THz) frequencies, the relevant dynamics and hence the observation time interval $\tau$ are both below the picoseconds time-scale and the previous assumption of Markovianity, i.e.,  $t_D \ll \tau$, starts to break down. Under the condition $t_D \sim  \tau$, non-Markovian stochastic Schr\"odinger equations have been proposed which allow an alternative procedure for solving the reduced state $|\psi_{q}(t)\rangle$~\cite{diosi1998non,strunz1999open,gambetta2002non,gambetta2002perturbative,budini2000non,bassi2002dynamical,bassi2003stochastic}. 
However, non-Markovian SSEs constructed from Eq.~\eqref{cond}, unlike the Markovian SSEs, suffer from interpretation issues~\cite{gambetta2002non}. 
In the non-Markovian regime, the perturbation of the environment due to the quantum backaction of a measurement at time $t$ would not be washed out in the time lapse $\tau \sim t_D$ and hence the joint probability distribution $P(q_t)P(q_{t+\tau})$ would not become separable, i.e. $P(q_t,q_{t+\tau})) \neq P(q_t)P(q_{t+\tau})$. As a direct consequence, connecting in time the different solutions $q_t$ and $q_{t+\tau}$ (sampled independently from the probability distributions $P(q_t,t)$ and $P(q_{t+\tau},{t+\tau})$ as in Figure~\ref{mark} to make a trajectory ``would be a fiction''~\cite{gambetta2002non,diosi2008non,PhysRevLett.101.140401}. 
Here, the word ``fiction'' means that the time-correlations computed in Equation~\eqref{corr_funct_markov} are wrong, i.e., the expectation value in Equation~\eqref{corr_funct_markov} would simply be different from the experimental result.


According to D'Espagnat the above discussion can be rephrased in terms of the so-called proper and improper mixtures~\cite{d2018conceptual} . 
Following D'Espagnat arguments, the reduced density matrix in Equation~\eqref{result} is an improper mixture because it has been constructed by tracing out the degrees of freedom of the environment. On the contrary, a proper mixture is a density matrix constructed to simultaneously define several experiments where a closed system is described, at each experiment, by different pure states. 
Due to our ignorance, we do not know which pure state corresponds to which experiment, so we only know the probabilities of finding a given pure state. D'Espagnat argues that the ignorance interpreation of the proper density matrix, cannot be applied in the improper desity matrix discussed here (See Appendix  \ref{espagnat}). 
To understand why under a Markovian regime open systems can be described by pure states (using a proper mixture), we remind that Markovianity implies conditions on the observation time. For a given correlation time $t_D$, a given open system can be in the Markovian or non-Markovian regimes depending on the time of observation $\tau$. That is, for small enough observation times all open systems are non-Markovian and hence must be understood as an improper mixture. On the contrary, for large enough observation times,  open systems can behave as closed systems (with a negligible entanglement with the environment) and be effectively represented by pure states.

\subsection{The Bohmian interpretation of conditional states}
\label{Bohm_interpretation}

So, under non-Markovian (i.e., the most general) conditions, the conditional states $|\psi_{q}(t) \rangle$ can be used to reconstruct the reduced density matrix as in Equation~\eqref{result} but cannot be required to provide further information by its own. 
This interpretation problem is rooted on the fact that orthodox quantum mechanics does only provide reality to objects whose properties (such as $q$) are being directly measured. But, as explained in the previous subsection, it is precisely the fact of introducing the measurement of $q$ (without including the pertinent backaction on the system evolution) that prevents the conditional states $|\psi_{q}(t) \rangle$ of the non-Markovian SSE to be connected in time for $t_D\sim\tau$. 
In this context, a valid question for the interpretation of $|\psi_{q}(t) \rangle$ is whether or not we can obtain information of, e.g., the observable $Q$ without perturbing the state of the system. The answer given by orthodox quantum mechanics is crystal clear: except for Markovian conditions this is not possible because information requires a measurement, and the measurement induces a perturbation. Notice, however, that the assumption that only measured properties are real is not something forced on us by experimental facts, but it is a deliberate choice of the orthodox quantum theory. 
Therefore we here turn to a nonorthodox approach: the Bohmian interpretation of quantum mechanics~\cite{benseny2014applied,pladevall2019applied,holland1995quantum,bohm1952suggested,durr2004bohmian,durr1992quantum}.

A fundamental aspect of the Bohmian theory is that reality (of the properties) of quantum objects does not depend on the measurement. That is, the values of some observables, e.g., the value of the positions of the particles of the environment, exist independently of the measurement. If $q$ is the collective degree of freedom of the position of the particles of the environment and $x$ is the collective degree of freedom of the position of particles of the system, then, the Bohmian theory defines an experiment in the laboratory by means of two basic elements: (i) the wavefunction $\langle q,x |\Psi(t) \rangle=\Psi(x,q,t)$ and (ii) an ensemble of trajectories $Q^i(t),X^i(t)$ of the environment and of the system.  We use a superindex $i$ to denote that each time an experiment is repeated, with the same preparation for the wavefunction $\Psi(x,q,t)$, the initial  positions of the environment and system particles can be different. They are selected according to the probability distribution $|\Psi(X^i,Q^i,0)|^2$~\cite{pladevall2019applied}. 
The equation of motion for the wavefunction $\Psi(x,q,t)$ is the time-dependent Schr\"odinger equation in Equation~\eqref{schr}, while the equations of motion for the environment and system trajectories $Q^i(t),X^i(t)$ are obtained by time-integrating the velocity fields $v_q(x,q,t)=J_q(x,q,t)/|\Psi(x,q,t)|^2$ and $v_x(x,q,t)=J_x(x,q,t)/|\Psi(x,q,t)|^2$ respectively. Here, $J_q(x,q,t)$ and $J_x(x,q,t)$ are the standard current densities of the environment and the system respectively \cite{footnote1}. According to the continuity equation
\begin{equation}\label{continutiy}
  \frac{d |\Psi(x,q,t)|^2}{dt}+{\nabla_x} (v_x(x,q,t)|\Psi(x,q,t)|^2)+{\nabla_q} (v_q(x,q,t)|\Psi(x,q,t)|^2)=0,
\end{equation}
the ensemble of trajectories $\{Q(t),X(t)\}=\{Q^1(t),X^1(t),Q^2(t),X^2(t)....Q^M(t),X^M(t)\}$ with $M\to \infty$ can be used to reproduce the probability distribution $|\Psi(x,q,t)|^2$ at any time. Thus, by construction, the computation of ensemble values from the orthodox and Bohmian theories are fully equivalent, at the empirical level.

From the full wavefunction $\langle x,q |\Psi(t) \rangle=\Psi(x,q,t)$ (solution of Equation~\eqref{Scho}) and the trajectories $Q^i(t),X^i(t)$, one can then easily construct the Bohmian conditional wavefunction of the system and environment as  
  $\tilde \psi_{Q^i(t)}(x,t)=\Psi(x,Q^i(t),t)$,
and
  $\tilde \psi_{X^i(t)}(q,t)=\Psi(X^i(t),q,t)$
respectively.
Notice that this Bohmian definition of conditional states
does not require to specify if the system is measured or not because of the ontological nature of the trajectories $\{Q(t),X(t)\}$. Consequetly, the conditional wavefunctions $\tilde \psi_{Q^i(t)}(x,t)$ contain all the required information to evaluate dynamical properties of the system no matter whether Markovian or non-Markovian conditions are being considered. This can be seen by noticing that the velocity of the trajectory $X^i(t)$ given by $v_q(X^i(t),Q^i(t))$ can be equivalently computed either from (the $x-$spatial derivatives of) the global wavefunction $\Psi(x,Q,t)$ evaluated at $X^i(t)$ and $Q^i(t)$, or from (the $x$-spatial derivative of) the conditional wavefunction $\tilde \psi_{Q^i(t)}(x,t)$ evaluated at $X^i(t)$ (both velocities are identical). Thus, in a particular experiment $i$ and for a given time $t$, the dynamics of the Bohmian trajectory $X^i(t)$ can be computed either from $\tilde \psi_{Q^i(t)}(x,t)$ or from $\Psi(x,q,t)$.  

The Bohmian conditional wavefunction of the system can now be connected to the Orthodox conditional wavefunction in Equation~\eqref{cond} by imposing $Q^i(t)=q_t$. Then one can readily write:
\begin{equation}\label{B_cond_s}
    |\tilde\psi_{q_t}(t)\rangle = P(q_t,t)|\psi_{q_t}(t)\rangle.
\end{equation}
At first sight, one can think that the difference between the Bohmian and orthodox conditional states is just a simple renormalization constant $P(q_t,t)$ (see Appendix \ref{mean_boh} for a more detailed explanation of the role of this renormalization constant). However, the identity in Equation~\eqref{B_cond_s} has to be understood to be satisfied at any time $t$ which also implies that the identity $Q^i(t)=q_t$ should be satisfied at any time. If we describe another experiment $Q^j(t)=q_{t}'$ , we have to define another conditional state $|\tilde\psi_{q_{t}'}(t)\rangle$ and so It can be possible that, at one particular time $t=t_1$, both conditional states are identical i.e. $|\tilde\psi_{q_{t_1}}(t_1)\rangle=|\tilde\psi_{q_{t_1}'}(t_1)\rangle$. However, this does not imply that both conditional wavefunctions identically describe the open system in one experiment. This is because every Bohmian trajectory has a fundamental role in describing the history of the Bohmian conditional state at one  particular experiment. Therefore, the trajectory $Q^i(t)$ uniquely describes the evolution of the conditional wavefunction $|\psi_{q_t}(t)\rangle$ at one experiment (the one labelled by the index $i$ in the Bohmian language) the same way as the trajectory $Q^j(t)$ and the conditional wave function $|\tilde\psi_{q_{t_1}'}(t_1)\rangle$ describes the experiment labelled by $j$. This comprehensive information about which particular experiment we are considering is missing in the orthodox counterpart of the conditional wavefunction, where the information of the environment $q_t$ is used by a particular system trajectory as opposed to the Bohmian where the same information of the environment can be used by different system Bohmian trajectories. As we said, $|\tilde\psi_{q_{t_1}}(t_1)\rangle=|\tilde\psi_{q_{t_1}'}(t_1)\rangle$ are the same orthodox conditional wave function, but not the same Bohmian conditional wavefunction. These differences explains precisely why SSEs cannot be connected in time and used to study the time-correlation of non-Markovian open system whereas the same can be done through the Bohmian conditional states, without any ambiguity.   

The mathematical definition of the measurement process in Bohmian mechanics and in the orthodox quantum mechanics differs substantially~\cite{pladevall2019applied}. In the orthodox theory a collapse (or reduction) law, different from the Schr\"odinger equation, is necessary to describe the measurement process~\cite{holland1995quantum}. Contrarily, in Bohmian mechanics the measurement is treated as any other interaction as far as the degrees of freedom of the measuring apparatus are taken into account~\cite{pladevall2019applied}. 
Therefore, while in the orthodox theory the conditional states $|\psi_{q_t}(t)\rangle$ cannot be understood without the perturbation of the full wavefunction $\Psi(x,q,t)$, in Bohmian mechanics the states $|\tilde\psi_{q_t}(t)\rangle$ do have a physical meaning even when the full wavefunction $\Psi(x,q,t)$ is unaffected by the measurement of the environment~\cite{gambetta2003interpretation}.
Interestingly, this introduces the possibility of defining what we call ``unmeasured (Bohmian) conditional states'' when it is assumed that there is no measurement or that the measurement of $q_t$ at time $t$ has a negligible influence on the subsequent evolution of the conditional state.  
\begin{figure}
\centering
\includegraphics[width=\textwidth]{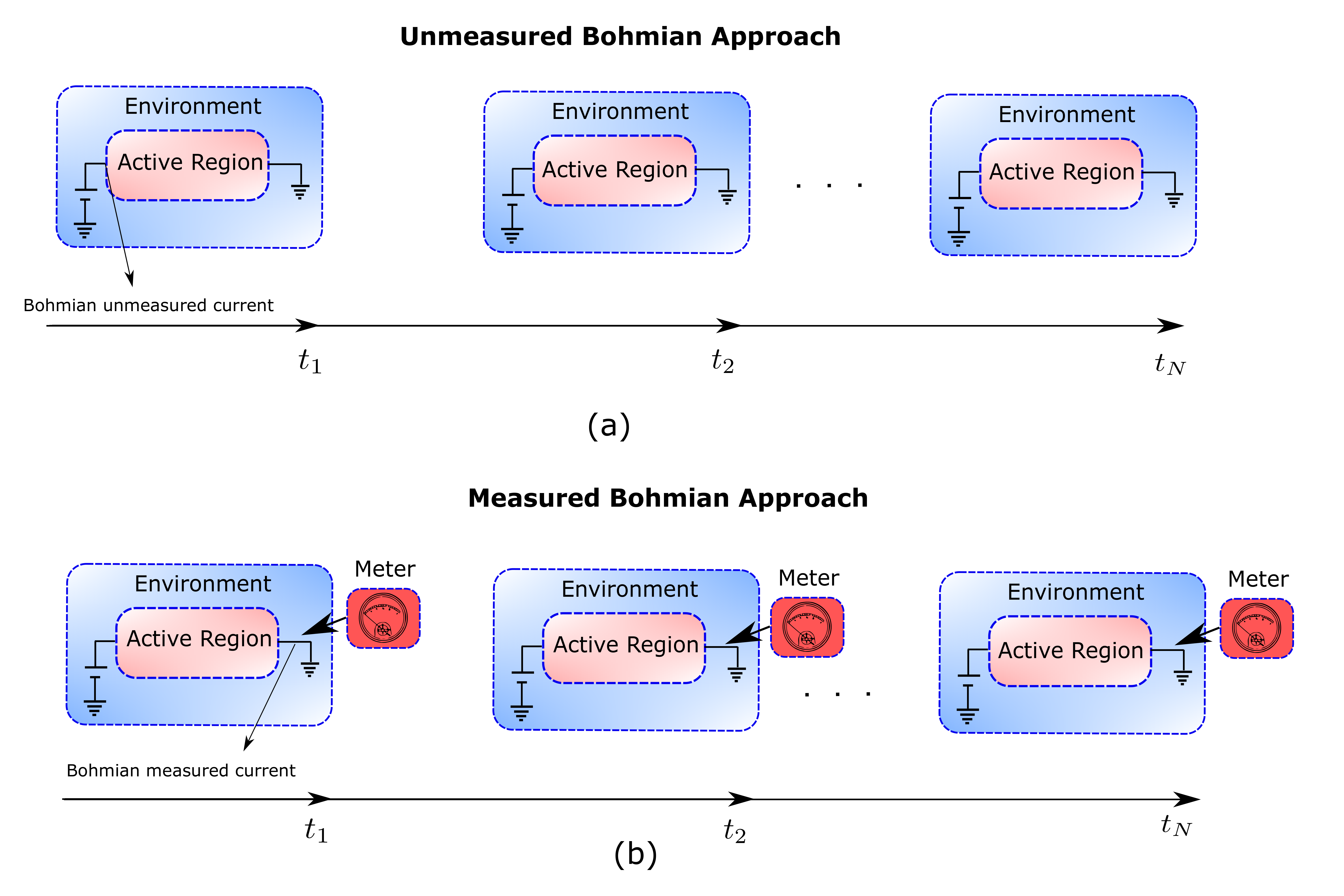}
\caption{(\textbf{a}) Figure depicting the Unmeasured Bohmian approach in which the computation of any property (electric current in an electron device) is independent of the measuring apparatus. (\textbf{b}) The continuous measurement of the electric current through an ammeter (measuring apparatus) can be also described in Bohmian mechanics by including the degrees of freedom of the measuring apparatus.}
\label{SSE_bisbis_fig}
\end{figure}

\begin{figure}
\centering
\includegraphics[width=0.5\textwidth]{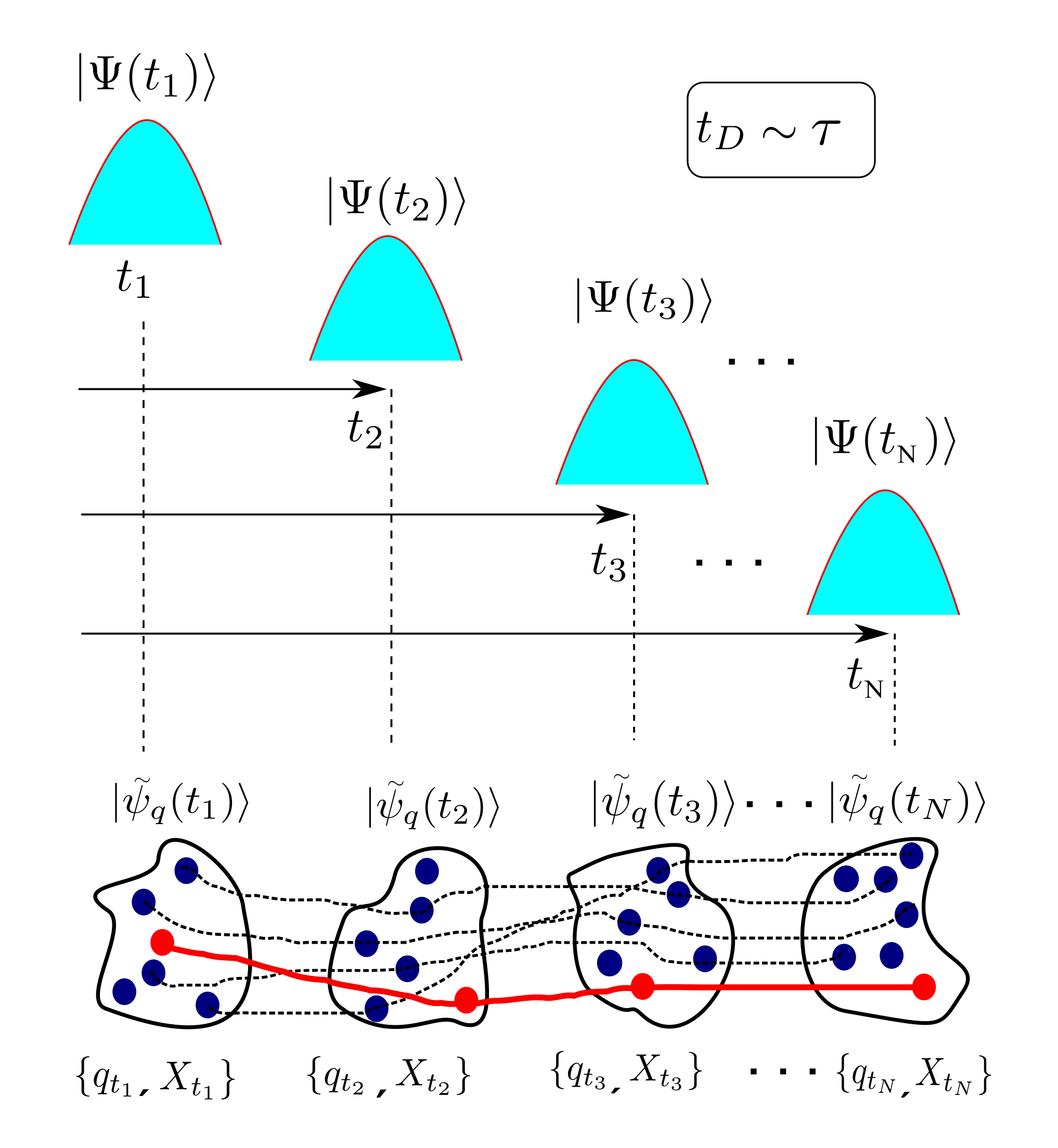}
\caption{Schematic representation of the combined system+environment wavefunction (blue gaussians) that is measured at different times and results in a Bohmian conditional sate $|\tilde\psi_{q}(t)\rangle$ conditioned to the set of environment values $\{q_t\}$ shown in dark blue circles. In the non-Markovian regime only those values from the set of values satisfying the continuity equation in Equation \eqref{continutiy} can be linked in time to form a trajectory (shown as connected red circles). Dashed lines represent connections that do not follow the continuity equation and hence cannot be used to evaluate any dynamics property.}
\label{mark_b}
\end{figure}

Importantly, the Bohmian conditional states can be used not only to reconstruct the reduced density matrix in Equation~\eqref{result} at any time, but the environment trajectories $\{Q(t)\}$ allow us to correctly predict any dynamic property of interest including time-correlation functions, e.g.:
\begin{equation}\label{corr_funct_nonmarkov}
  \langle Q(t) Q(t+\tau) \rangle= 
  \frac{1}{M} \sum_{i=1}^{M} Q^i(t) Q^i(t+\tau) = \int\int P(q_t,q_{t+\tau}) q_t q_{{t+\tau}} dq_t dq_{{t+\tau}},
  \end{equation}
where $M \to \infty$ is the number of experiments (Bohmian trajectories) considered in the ensemble and we have defined $P(q_t,q_{t+\tau})=\frac{1}{M} \sum_{i=1}^{M} \delta(q_t-Q^i(t)) \delta(q_{{t+\tau}}-Q^i({t+\tau}))$. 
As it is shown in Figure~\ref{mark_b}, the evaluation of Equation~\eqref{corr_funct_nonmarkov} and any other dynamic property when $t_D \sim \tau$ can be done only by connecting 
the (Bohmian) trajectories at different times in accordance with the continuity equation in Equation~\eqref{continutiy}.
This is in contrast with the evaluation of the dynamics in the Markovian regime where any position of the environment at time $t_1$ can be connected to another position of the environment at time $t_2$ (see Figure~\ref{mark}) and hence we can write $
  \langle Q(t) Q(t+\tau) \rangle \stackrel{t_D \ll \tau}{=} 
  \frac{1}{M^2} \sum_{i,j}^{M} Q^i(t) Q^j(t+\tau)$.
This very relevant point was first explained by Gambetta and Wiseman~\cite{gambetta2003interpretation,gambetta2003modal}. 


Although the Bohmian theory can also provide measured properties of the system that coincide with the orthodox results in Figure~\ref{SSE_fig}b, let us emphasize once more the merit of the unmeasured properties provided by the Bohmian theory, which remains mainly unnoticed in the literature. As it has been already explained, in the orthodox theory, measuring a particular value of the environment property $q$ at time $t$ cannot be conceived without the accompanying perturbation of the wavefunction $\Psi(x,q,t)$. Under non-Markovian conditions, it is precisely this perturbation that prevents the conditional states of the system $|\psi_{q_t}(t)\rangle$ to be connected in time to form a trajectory. Contrarily, in Bohmian mechanics, the existence of the environment trajectories $\{Q(t)\}$, even in the absence of any measurement, allows the possibility of connecting in time the conditional states $|\tilde\psi_{q_t}(t)\rangle$ even when $t_D \sim \tau$.

\section{Bohmian conditional wavefunction approach to quantum electron transport}
\label{BITLLES}

The different notions of reality invoked by the orthodox quantum theory and Bohmian mechanics lead to practical differences in the abilities that these theories can offer to provide information about quantum dynamics.   
Specifically, we have shown that contrarily to orthodox quantum mechanics, Bohmian mechanics allows to physically interpret (i.e., link in time) the conditional states of the SSE approach in general non-Markovian scenarios. The reason is that whereas in the Bohmian theory the reality of the current is independent of any measurement, the orthodox theory gives reality to the electrical current only when it is being measured (this is the so-called eigenstate-eigenvalue link). 
From the practical point of view this has a remarkable consequence. In the Bohmain approach the total current can be defined in terms of the dynamics of the electrons (Bohmian) trajectories without the need to define a measurement operator. As it will be shown in this section, the possibility of computing the total current at high frequencies without specifying the measurement operator is certainly a great advantage~\cite{pladevall2019applied}. In particular, one can then avoid cumbersome questions like, is the measurement operator strong or weak? If weak, how weak? How often do I need to measure? Every picosecond or every femtosecond? At high frequencies, how should I include the contribution of the displacement current in my current operator? 

In this section we provide a brief summary of the path that the authors of this work followed for developing an electron transport simulator based on the use of Bohmian conditional states. The resulting computational tool is called BITLLES~\cite{BITLLES0,BITLLES00,BITLLES000,BITLLES1, BITLLES2,BITLLES3, BITLLES5,BITLLES6,BITLLES7, BITLLES8}.  
Let us start by considering an arbitrary quantum system. The whole system, including the open system, the environment and the measuring apparatus, is described by a Hilbert space $\mathcal{H}$ that can be decomposed as $\mathcal{H}=\mathcal{H}_x\otimes\mathcal{H}_q$ where $\mathcal{H}_x$ is the Hilbert space of the open system and $\mathcal{H}_q$ the Hilbert space of the environment. If needed, the Hamiltonian $\mathcal{H}_q$ can include also the degrees of freedom of the measuring apparatus as explained in section~\ref{Bohm_interpretation}. We define  ${x}=\{{x}_{1},{x}_{2}...{x}_{n}\}$ as the degrees of freedom of $n$ electrons in the open system, while $q$ collectively defines the degrees of freedom of the environment (and possibly the measuring apparatus). The open system plus environment Hamiltonian can then be written as:
\begin{equation}
    \hat H = \hat{H}_{q}\otimes\hat I_\text{x}+\hat I_{q}\otimes\hat{H}_\text{x}+\hat V
\end{equation}
where $\hat{H}_\text{x}$ is the Hamiltonian of the system,  $\hat{H}_{q}$ is the Hamiltonian of the environment (including the apparatus if required), and $\hat V$ is the interaction Hamiltonian between the system and the environment. We note at this point that the number of electrons $n$ in the open system can change in time and so the size of the Hilbert spaces $\mathcal{H}_x$ and $\mathcal{H}_q$ can depend on time too.   

The equation of motion for the Bohmian conditional states $\langle x|\tilde\psi_{q_t}(t)\rangle=\tilde\psi_{q_t}(x,t)$ in the position representation of the system can be derived by projecting the many-body (system-environment) Schr\"odinger equation into a particular trajectory of the environment $q_t=Q(t)$, i.e.~\cite{PhysRevLett.98.066803,albareda2014correlated}: 
\begin{equation}\label{EQM_conditional} 
    i\hbar\frac{d\tilde\psi_{q_t}(x,t)}{dt} =  \langle q_t|\otimes\langle x|\hat{H} | \Psi( t)\rangle + i\hbar \nabla_q \langle q|\otimes\langle x|\Psi(t)\rangle\big{|}_{q=q_t} \frac{dq_t}{dt}.
\end{equation}
Equation~\eqref{EQM_conditional} can be rewritten as: 
\begin{equation}
    i\hbar\frac{d\tilde\psi_{q_t}(x,t)}{dt} =  \left[-\frac{\hbar ^2}{2m}\nabla^2_{x}+U^{eff}_{q_t}(x,t)\right] \tilde\psi_{q_t}(x,t)
    \label{EQM_conditional2}
\end{equation}
where 
\begin{equation}
\tilde U^{eff}_{q_t}(x, t)= {U}(x,t)+ V(x,q_t,t)+\mathcal{A}(x,q_t,t)+i\mathcal{B}(x,q_t,t).
\label{potentials}
\end{equation}
In Equation~\ref{potentials}, $U(x,t)$ is an external potential acting only on the system degrees of freedom, $V(x,q_t,t)=\langle q|\otimes \langle x|\hat V|\Psi\rangle/\Psi(x,q,t)\big|_{q=q_t}$ is the Coulomb potential between particles of the system and the environment evaluated at a given trajectory of the environment, $\mathcal{A}(x,q_t, t)=\frac{-\hbar^2}{2m}\nabla^2_{q}\Psi(x,q,t)/\Psi(x,q,t)\big|_{q=q_t}$ and $\mathcal{B}(x, q_t, t)=\hbar \nabla_{q}\Psi(x,q,t)/\Psi(x,q,t)\big|_{q=q_t} \dot{q}_t$ (with $\dot{q}_t = dq_t/dt$) are responsible for mediating the so-called kinetic and advective correlations between system and environment~\cite{albareda2014correlated,PhysRevLett.98.066803}. Equation~\eqref{EQM_conditional2} is non-linear and describes a non-unitary evolution.

Bohmian conditional states can be used to exactly decompose the unitary time-evolution of a closed quantum system in terms of a set of coupled, non-Hermitian, equations of motion~\cite{PhysRevLett.98.066803}. Inspired by the Bohmian trajectory based approach, conditional states allow one to describe the evolution of arbitrary subsets of the degrees of freedom in a system, on a formally exact level~\cite{PhysRevLett.98.066803,albareda2014correlated,norsen2015can,PhysRevMaterials.3.023803}.
An approximate solution of Equation~\eqref{EQM_conditional2} can always be achieved by making an educated guess for the terms $\mathcal{A}$ and $\mathcal{B}$ according to the problem at hand. For example the decoherence appearing due to the electrons interacting with the environment can be modeled by the electron-phonon interaction with the momentum change in the process encapsulated by the Hamiltonian describing the scattering of the electron with a phonon at a particular instant of time~\cite{colomes2017quantum}.
Specifically, in the BITLLES simulator the first and second terms in Equation~\eqref{potentials} are evaluated through the solution of the Poisson equation~\cite{BITLLES2}. The third and fourth terms are modeled by a proper injection model~\cite{zhen2} as well as proper boundary conditions~\cite{BITLLES1,BITLLES7,albareda2013self} that include the correlations between active region and reservoirs. Electron-phonon decoherence effects can be also effectively included in Equation~\eqref{EQM_conditional2}~\cite{colomes2017quantum}.

In an electron device, the number of electrons contributing to the electrical current are mainly those in the active region of the device. The number fluctuates as there are electrons entering and leaving the active region. Thus it is necessary to somehow model the addition and subtraction of the electrons in the active region. This creation and destruction of electrons leads to an abrupt change in the degrees of freedom of the many body wavefunction which cannot be treated with a Schr\"{o}dinger-like  equation for $\tilde\psi_{q_t}(x, t)$ with a fixed number of degrees of freedom. 
In the Bohmian conditional approach, this problem can be circumvented by decoupling the system conditional wavefunction $\tilde\psi_{q_t}(x, t)$ into a set of conditional wavefunctions for each electron. 
More specifically, for each electron $x_i$, we define a single particle conditional wavefunction $\tilde{\tilde\psi}_{q_t}(x_i,\bar X_i(t),t)$, where $\bar X_i(t) = \{X_1(t),..,x_{i-1}(t),x_{i+1},..,X_n(t)\}$ are the Bohmian positions of all electrons in the active region except $x_i$, and the second tilde denotes the single-electron conditional decomposition that we have considered on top of the conditional decomposition of the system-environment wavefunction. The set of equations of motion of the resulting $n(t)$ single-electron conditional wavefunctions inside the active region can be written as:
\begin{eqnarray}\label{EQM_conditional5}
    i\hbar\frac{d\tilde{\tilde\psi}_{q_t}(x_1,\bar X_1(t),t)}{dt}& =&  \left[-\frac{\hbar ^2}{2m}\nabla^2_{x_1}+\tilde{\tilde U}^{eff}_{q_t}(x_1,\bar X_1(t),t)\right] \tilde{\tilde\psi}_{q_t}(x_1,\bar X_1(t),t)\\
    &\vdots&    \nonumber\\
    i\hbar\frac{d\tilde{\tilde\psi}_{q_t}(x_n,\bar X_n(t),t)}{dt} &=&  \left[-\frac{\hbar ^2}{2m}\nabla^2_{x_n}+\tilde{\tilde U}^{eff}_{q_t}(x_n,\bar X_n(t),t)\right] \tilde{\tilde\psi}_{q_t}(x_n,\bar X_n(t),t)
\end{eqnarray}
That is, the first conditional process is over the environment degrees of freedom and the second conditional process is over the rest of electrons on the (open) system. 

We remind here that the active region of an electron device, acting as the open system, is connected to the ammeter (acting as the measuring apparatus), by a macroscopic cable (acting as the environment) in a type of scenario given by Figure~\ref{SSE_fig}b. The electrical current provided by the ammeter is the relevant variable outside the system that we are interested in. At THz frequencies, however, the electrical current is not only the particle current, but also the displacement current.  It is well-known that the total current defined as the particle current plus the displacement current is a divergence-less vector~\cite{eisenberg2017dynamics,oriols2013quantum}. Consequently, the total current evaluated at the end of the active region is equal to the total current evaluated at the cables. So the variable of the environment associated to the total current, $q_t = I(t)$, can be equivalently computed at the borders of the open system. The reader is referred to Ref.~\cite{marian} for a discussion on how the $I(t)$ can be defined in terms of Bohmian positions with the help of a quantum version of the Ramo-Schokley-Pellegrini theorem~\cite{albareda2012computation}. In particular, it can be shown that the total (particle plus displacement) current in a two-terminal devices can be written as~\cite{albareda2012computation}:
\begin{equation}\label{ramo}
  I(t)=  \frac{e}{L} \sum_{i=1}^{n(t)} v_{x_i}(X_i(t),\bar X_i(t),t) = \frac{e}{L} \sum_{i=1}^{n(t)} \text{Im}\left( \frac{\nabla_{x_i} \tilde{\tilde\psi}_{q_t}(x_i,\bar X_i(t),t)}{\tilde{\tilde\psi}_{q_t}(x_i,\bar X_i(t),t)} \right)\Bigg{|}_{x_i = X_i(t)},
 \end{equation}
 where $L$ is the distance between the two (metallic) contacts, $e$ is the electron charge and $v_{x_i}(X_i(t),\bar X_i(t),t)$ is the Bohmian velocity of the $i$-th electron inside the active region. Let us note that  $I(t)$ is the electrical current given by the ammeter (although computed by the electrons inside the open system). Since the cable has macroscopic dimensions, it can be shown that the measured current at the cables is just equal to the unmeasured current (taking into account only the simulation of electrons inside the active region) plus a source of (nearly white) noise which is only relevant at very high frequencies~\cite{marian}. The basic argument is that the (non-simulated) electrons in the metallic cables have a very short screening time. The electric field of an electron in the cable goes to zero rapidly due to the presence of many other mobile charge carriers in the cable so that this outer electron has a negligible contribution to the displacement current evaluated at the border of the active region~\cite{pandey2019proposal}. 

\section{Numerical Results}
\label{nm}

In this section we present numerical results that were obtained with the BITLLES simulator (the methodology of which has been explained in the previous section) and that demonstrate its ability to provide dynamic information for both Markovian and non-Markovian scenarios. 
We simulate a two-terminal electron device whose active region is a graphene sheet contacted to the outer by two (ohmic) contacts. Graphene is a $2D$ material that has attracted a lot of attention recently because of its high electron mobility. It is a gapless material with linear energy band, which differs from the parabolic energy bands of traditional semiconductors. In graphene, the conduction and valence bands coincide at an energy point known as the Dirac point. Thus, the dynamics of electrons is no longer governed by an (effective mass) Schr\"{o}dinger equation, but by the Dirac equation, allowing transport from the valence to the conduction band (and vice versa) through Klein tunneling. A Bohmian conditional bispinor (instead of a conditional scalar wavefunction) is used to describe electrons inside the device. The change from a wavefunction to a bispinor does not imply any conceptual difficulty but just a mere increment of the computational cost. More details can be found in Appendix~\ref{graph}. 

\begin{figure}
\centering
\includegraphics[width=\textwidth]{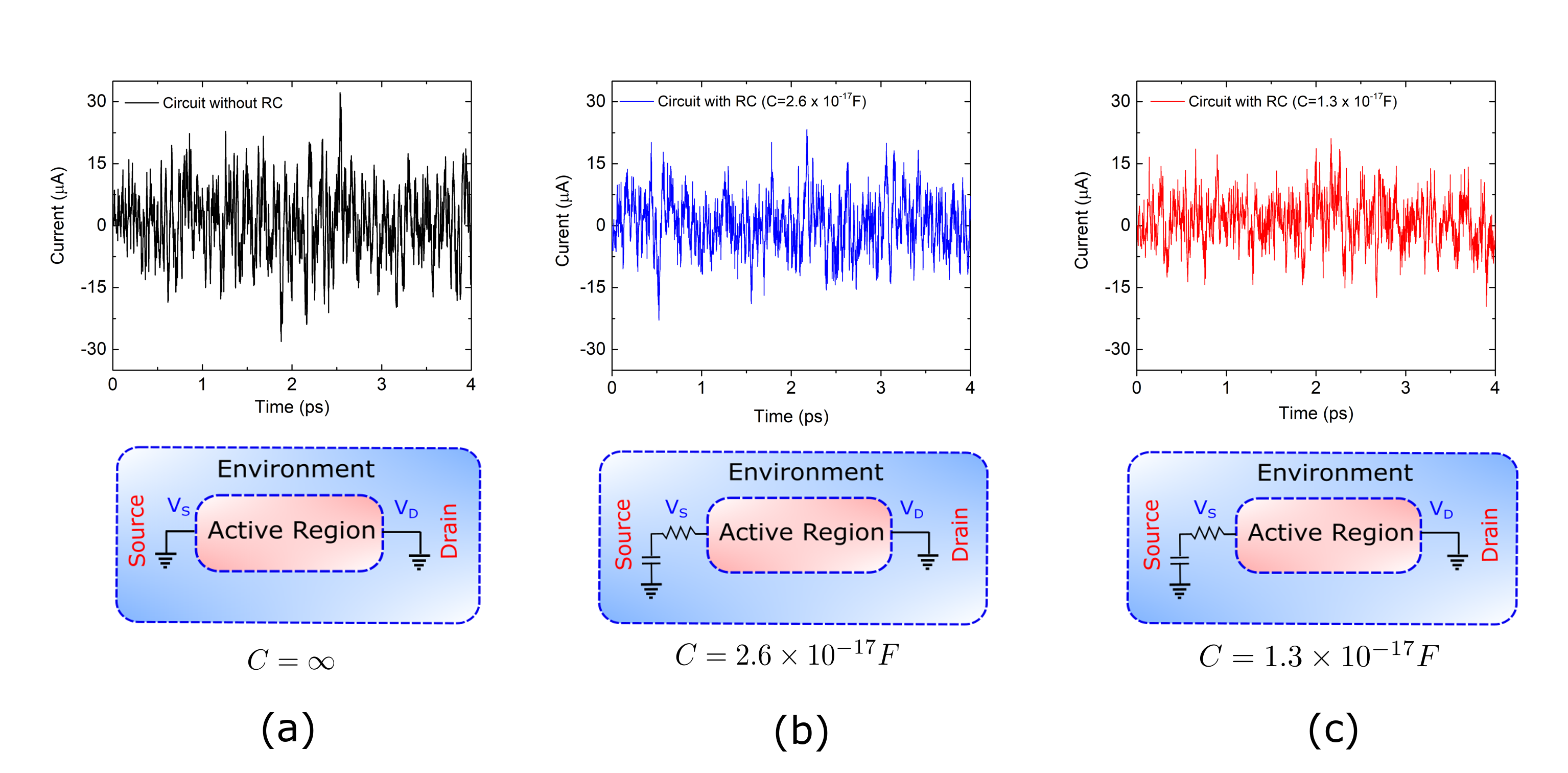}
\caption{Total (particle plus displacement) electrical current $I(t)$ evaluated at the ammeter as a function of time for a graphene device connected to three different RC circuits with $R=187\Omega$. The values of the capacitances are: (\textbf{a}) $C=\infty$, (\textbf{b}) $C = 2.6\times 10^{-17}$F and (\textbf{c}) $C = 1.3 \times 10^{-17}$F}
\label{Res1}
\end{figure}

In particular, we want to simulate electron transport in graphene at very high frequencies (THz) taking into account the electromagnetic environment of the electron device. Typically, nanoscale devices are small enough to assume that, even at THz frequencies, the electric field is much more relevant than the magnetic field. Therefore, only the Gauss law (first Maxwell's equations) is enforced to be fulfilled in a self-consistent way (i.e. taking into account the actual charge distribution in the active region). However, the environment of nanoscale devices is commonly (nearly) metallic and of macroscopic dimensions. In there, the magnetic and electric fields become both relevant, acting as active (detecting or emitting) THz antennas. For the typical electromagnetic modes propagating in the metals, the magnetic and electric fields are translated into the language of currents and voltages and the whole antenna is modelled as a part of an electric circuit. In this work, the graphene device interacts with an environment that is modelled by a Resistor (R) and a capacitor (C) connected in series through ideal cables (see the schematic plots in Figure \ref{Res1}a, \ref{Res1}b and \ref{Res1}c).  

The active region of the graphene device is simulated with the Bohmian conditional wavefunction approach explained in the previous section, while the RC circuit is simulated using a time-dependent finite-difference method. We consider the system plus environment to be in equilibrium. Specifically, the self-consistent procedure to get the current is as follows: an initial (at time $t=0$) zero voltage is applied at the source ($V_S(0)=0$) and drain ($V_D(0)=0$) contacts of the graphene active region. At room temperature this situation yields a non-zero current from Equation~\eqref{ramo} (i.e., $I(0)\neq 0$) because of thermal noise. Such current $I(0)$ enters the RC circuit and leads to a new voltage $V_S(dt)\neq 0$ at the next time step $dt$ (where $dt$ represents the time step that defines the interaction between the RC circuit and the quantum device which was set to $dt=0.5$fs). The new source $V_S(dt)\neq 0$ and fixed drain $V_D(dt)=0$ voltages now lead to a new value of the current $I(dt)\neq 0$ in~\eqref{ramo} which is different from zero not only because of thermal noise but also because there is now a net bias ($V_D(dt)-V_S(dt) \neq 0$). This new current $I(dt)$ is used (in the RC circuit) to get a new $V_S(2dt)$ that is introduced back in the  device to obtain $I(2dt)$ and so on so forth. Importantly, as the system and environment are in equilibrium, the expectation value of $I(t)$ is zero at any time, i.e.:
$\langle I(t) \rangle = 0 \;\; \forall t$.

We consider three different environments (with different values of the capacitance). In Figure~\ref{Res1}a we plot the total (particle plus displacement) electrical current at the end of the active region when $R=0$ and $C=\infty$. The same information is shown in Figures~\ref{Res1}b and \ref{Res1}c for two different values of the capacitance $C = 2.6\times 10^{-17}$F and  $C = 1.3\times 10^{-17}$F. In all cases the value of the resistance is $R=187\Omega$, and we assumed the current $I(t)$ to be positive when it goes from drain to source. 
The effect of the RC circuit is, mainly, to attenuate the current fluctuations, which are originated due to thermal noise. This can be seen by comparing Figure~\ref{Res1}a with Figures~\ref{Res1}b and \ref{Res1}c. The smaller the capacitance the smaller the current fluctuations. This can be explained as follows: when the net current is positive, the capacitor in the source starts to be charged and so the voltage at the source increases trying to counteract the initially positive current. Therefore, the smaller the capacitance the faster the RC circuit reacts to a charge imbalance. 
\begin{figure}
\centering
\includegraphics[width=\textwidth]{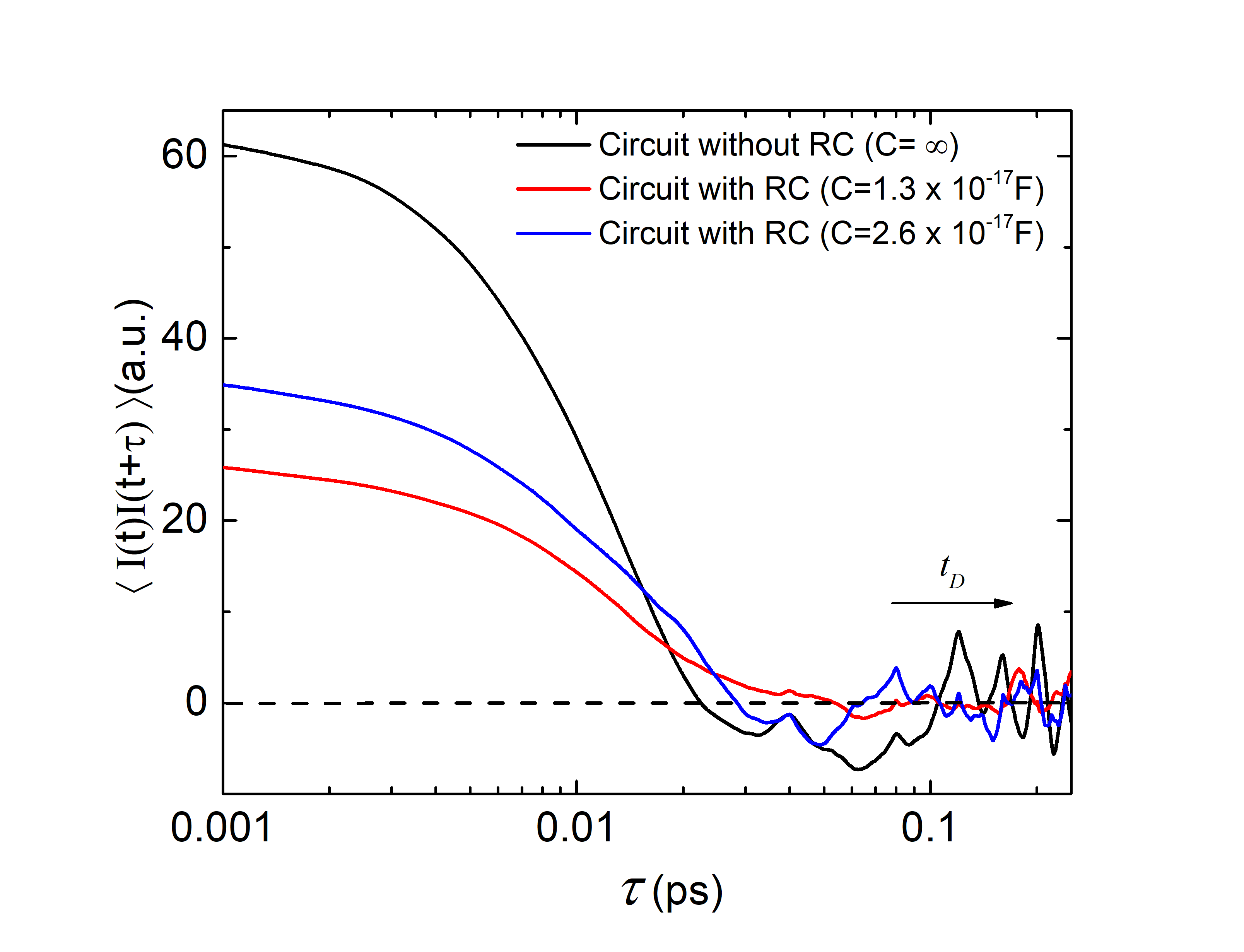}
\caption{Total current-current correlation as a function of time for the three different experiments in Figure~\ref{Res1}. The zero is indicated by a dashed line to show the tendency of the total current, understood as a property of the environment, to vanish at long times $\tau$. Zero autocorrrelation implies an independence between $I(t)$ and $I(t+\tau)$ which is typical for Markovian scenarios. This is not true for the short $\tau$ considered here which are the representatives of the non-Markovian dynamics.}
\label{Res2}
\end{figure}

In Figure~\ref{Res2} we plot the total (particle plus displacement) current-current correlations as a function of the observation time $\tau$ for the three scenarios in Figure~\ref{Res1}. Correlations at very small observation times provide information of the variance of the current, which, as explained above, is reduced as the value of the capacitance is increased. Numerical simulations (not shown here) exhibit that the role of the resistor $R$ is less evident because the active region itself has a much larger (than $R=187\Omega$) associated resistance. Numerically the distinction between Markovian and non-Markovian dynamics boils down to the comparison of time correlations as defined in Equations~\eqref{corr_funct_markov} and \eqref{corr_funct_nonmarkov}. Since there is no net bias applied to the graphene device (i.e., it is in equilibrium), an ensemble average of the current (over an infinite set of trajectories like the one depicted in Figure~\ref{Res1}) yields $\langle I(t) \rangle=0 \;\; \forall t$. Time correlation functions computed in Equation~\eqref{corr_funct_markov} are thus zero by construction, i.e.: 
    $\langle I(t)\rangle \langle I(t+\tau) \rangle=0  \;\; \forall t,\tau$.
Therefore, Figure~\ref{Res2} expressly shows the non-Markovian dynamics that occur at very high-frequencies (below the ps time-scale) and sets the correlation time of the environment at $t_D \sim \;$ps (i.e., $\langle I(t) I(t+\tau) \rangle = \langle I(t)\rangle \langle I(t+\tau) \rangle = 0 \;\; \forall \tau> 1\text{ps} \sim t_D$). 
Although all three values of the capacitance $C$ in Figure~\ref{Res2} yield the same order of magnitude for $t_D \sim \;$ps, it seems also true that the smaller the value of the capacitance, the smaller $t_D$. 

Current-current correlations shown in Figure~\ref{Res2} can be better understood by assessing the transit time of electrons. For a velocity of roughly $10^6$ m/s inside an active region of $L=40$nm length is roughly $\tau_T=L/v_x=0.04$ps. Positive correlations correspond to transmitted electrons travelling from drain to source (as well as electrons traversing the device from source to drain). During $0<t<\tau_T$ electrons are transiting inside the active region, such electrons provide always a positive (or negative) current as seen in expression (\ref{ramo}). In other words, if we have a positive current at time $t$ because electrons are travelling from drain to source, we can expect also a positive current at times $t'$ satisfying $t<t'<t+\tau_T$. The negative correlations belong to electrons that are being reflected. They enter in the active region with a positive (negative) velocity and, after some time $\tau_R$ inside the device, they are reflected and have negative (positive) velocities until they leave the device after spending roughly $2 \tau_R$ in the active region. Thus, during the time $\tau_R<t<2 \tau_R$  which will be different for each electron depending on the time when they are reflected, we can expect negative correlations.  Interestingly, during the $4$ ps simulation the number of Bohmian trajectories reflected are the double in the black ($C=\infty$) simulation than in the red one ($C=1.3 \times 10^{17}$ F). This can be explained in a similar way as we explained the reduction of the current fluctuations. The  fluctuations of the electrical current imply also fluctuations of the charge inside the active region, which are translated (through the Gauss law) into fluctuations of the potential profile. Thus, the larger noisy current, the larger noisy internal potential profile. This implies a larger probability of being reflected by the Klein tunneling phenomenon.
Therefore, if one aims at describing the dynamics of nanoscale devices with a time-resolution $\tau$ that is comparable to (or goes beyond) the electron transit time $\tau_T$, a non-Markovian approach is necessary. And this is so because the total current $I(t)$ (which has contributions from the displacement and the particle currents) shows correlations at times that are smaller than the electron transit time.

\section{Conclusions and final remarks}

\label{conclusion}

Theoretical approaches to open quantum systems that rely on the manipulation of state vectors instead of a reduced density matrix have well known computational advantages. Two major benefits are the substantial reduction of the dimensionality of the involved mathematical objects and the preservation of complete positivity~\cite{rivas2014quantum}.
But, substituting density matrices by state vectors constitutes also an attempt to achieve a more detailed description of the dynamics of open quantum systems~\cite{breuer2002theory,diosi2008non}. 
It is well recognized, for example, that the continuous measurement of an open quantum system with associated Markovian dynamics can be described by means of a stochastic Schr\"odinger equation (see Table~\ref{tab}.O4). The conditional state solution to such an equation over some time interval can be linked to a ``quantum trajectory''~\cite{carmichael2009open,diosi2008non} of one property of the environment. Thus, the conditional state can be interpreted as the state of the open system evolving while its environment is under continuous monitoring. This is true in general for Markovian systems, no matter whether or not the environment is being actually measured (i.e., it is valid for both Figures~\ref{schr}a and \ref{schr}b). This fact, is of great importance for designing and experimentally implementing feedback control in open quantum systems~\cite{wiseman2009quantum}. If this interpretation could also be applied to non-Markovian SSEs~\cite{diosi1998non,strunz1999open}, then this would be very significant for quantum technologies, especially in condensed matter environments (e.g., electron devices), which are typically non-Markovian~\cite{breuer2002theory}.

Unfortunately, for non-Markovian conditions, the above interpretation is only possible for the rather exotic scenario where the environment is being continuously monitored and the system is strongly coupled to it. 
As no correlation between the system and the environment can build up, the evolved system is kept in a pure state. This is the well-known quantum Zeno regime~\cite{misra1977zeno,home1997conceptual}, under which conditional states can be trivially used to describe the frozen properties of the system (see Table~\ref{tab}.O1). 
Without the explicit consideration of the measurement process (as in Figure~\ref{schr}a), however, the postulates of the orthodox theory restrict the amount of dynamical information that can be extracted from state vectors (see Table~\ref{tab}.O2). 
In most general conditions, for $\tau>0$ and non-Markovian dynamics, while conditional states can be used to reconstruct the reduced density matrix, they cannot be used to evaluate time-correlations (see Table~\ref{tab}.O3)~\cite{gambetta2003interpretation,wiseman2008pure}. And this is not only true when the environment is being measured (as in Figure~\ref{schr}b), but also when it is not measured (as in Figure~\ref{schr}a).  

Therefore we turned into a nonorthodox approach: the Bohmian interpretation of quantum mechanics. The basic element of the Bohmian theory (as in many other quantum theories without observers) is that the intrinsic properties of quantum systems do not depend on whether the system is being measured or not. Such ontological change is, nevertheless, fully compatible with the predictions of orthodox quantum mechanics because the reality of quantum objects is not in contradiction with non-local and contextual phenomena. 
And yet, the ontological nature of the trajectories in Bohmian mechanics introduces the possibility of evaluating dynamic properties in terms of conditional wavefunctions for Markovian and non-Markovian dynamics, no matter whether the environment is being actually measured or not (see Table~\ref{tab}.B1-B4 and Figures~\ref{SSE_bisbis_fig}a and \ref{SSE_bisbis_fig}b).  
\begin{table}
\centering
\begin{tabular}{|c|c|c|c|c|}
\hline
\thead{Validity of conditional\\ 
states to provide \\ dynamic information} & {\thead{Non-Markovian \\-measured-\\$t_D>\tau = 0$}} &  {\thead{Non-Markovian \\-unmeasured-\\$t_D>\tau = 0$}} & {\thead{Non-Markovian \\ -(un)measured-\\$t_D \sim \tau > 0$}} & {\thead{Markovian\\-(un)measured-\\$t_D \ll \tau$}}                   \\ \hline
                  \textbf{Orthodox} &  (O1) \cmark     &  (O2) \xmark &  (O3) \xmark    & (O4) \cmark       \\ \hline
                 \textbf{Bohmian} & (B1) \cmark  & (B2) \cmark  & (B3) \cmark   & (B4) \cmark      \\ \hline
\end{tabular}%
\caption{Validity of Bohmian vs orthodox conditional states  to provide dynamic information of open quantum system depending on the relation between the environment decoherence time $t_D$ and the observation period $\tau$. Here (un)measured refers to unmeasured and measured indistinctively.}
\label{tab}
\end{table}

In summary, the Bohmian conditional states lend themselves as a rigorous theoretical tool to evaluate static and dynamic properties of open quantum systems in terms of state vectors without the need of reconstructing a reduced density matrix. The price to be paid is that for developing a SSE-like approach based on Bohmian mechanics, one needs to compute both the trajectories of the environment and of the system. 
Let us also notice that here we have always assumed that the positions of the environment are the variables that the states of the system are conditioned to. However, it can be shown that the mathematical equivalence of the SSEs with state vectors conditioned to other "beables" of the environment (different from the positions) is also possible. It requires using a  generalized modal interpretation of quantum phenomena, instead of the Bohmian theory. A review on the modal interpretation can be found in~\cite{colodny1972paradigms, cabello2016thermodynamical}. 

As an example of the practical utility of the Bohmian conditional states, we have introduced a time-dependent quantum Monte Carlo algorithm, called BITLLES, to describe electron transport in open quantum systems. We have simulated a graphene electron device coupled to an RC circuit and computed its current-current correlations up to the THz regime where non-Markovian effects are relevant. The resulting simulation technique allows to describe not only DC and AC device's characteristics, but also noise and fluctuations. Therefore, BITLLES extends to the quantum regime the computational capabilities that the Monte Carlo solution of the Boltzmann transport equation has been offering for decades for semi-classical devices.


\acknowledgments{The authors are grateful to Howard Wiseman and Bassano Vacchini for enlightening comments. We acknowledge also financial support from Spain's Ministerio de Ciencia, Innovación y Universidades under Grant No. RTI2018-097876-B-C21 (MCIU/AEI/FEDER, UE). the European Union's Horizon 2020 research and innovation programme under grant agreement No Graphene Core2 785219 and under the Marie Skodowska-Curie grant agreement No 765426 (TeraApps). G.A. also acknowledges financial support from the European Unions Horizon 2020 research and innovation programme under the Marie Skodowska-Curie Grant Agreement No. 752822, the Spanish Ministerio de Economa y Competitividad (Project No. CTQ2016-76423-P), and the Generalitat de Catalunya (Project No. 2017 SGR 348).}

\appendix

\section{D'Espagnat distinction between "proper" and "improper" mixtures}
\label{espagnat}

An alternative explanation on the difficulties of state vectors in  describing open quantum system comes from the distinction between "proper" and "improper" mixtures by D’Espagnat. 

\begin{itemize}
\item The "proper" mixture is simply a mixture of different pure states of a closed system. We define such pure states as $|\psi_q\rangle$ with $q=1,..,N$. We know that each of these states represent the quantum nature of the closed system in one of the repeated experiments,  but we ignore which state corresponds to each experiment. We only know the probability  $P(q_t)$ that one experiment is represented by the pure state $|\psi_q\rangle$. Then, if we are interested in computing some ensemble value of the system, over all experiments, von Neumann introduced the mixture  $\rho=\int p_q |\psi_q\rangle \langle \psi_q| dq$. It is important to notice that we are discussing here human ignorance (not quantum uncertainty). The system is always in a well-defined state (for all physical computations), but we (the humans) ignore which is the state in each experiment.  

\item The "improper" mixture refers to the density matrix that results from a trace reduction of a pure sate (or statistical operator) of a whole system that includes the system and the environment. The reduced density of the system alone is given by tracing out the degrees of freedom of the environment, giving the result in Equation~\eqref{result}, which is mathematically (but not physically) equivalent to the results of the "proper" mixture constructed from our "ignorance" of which state represents the system.   
\end{itemize}

D’Espagnat claims that the ignorance interpretation of the "proper" mixture cannot be given to the "improper" mixture. The D'Espagnat's argument is as follows. Let us assume a pure global system (inclding the open system and the environment) described by Equation~\eqref{Scho}.Then,  if we accept that the physical state of the system is given by $|\psi_{q_t}(t)\rangle$, we have to accept that the system-plus-environment is in the physical state $|q\rangle \otimes |\psi_{q_t}(t)\rangle$ with probability $P(q_t)$. The ignorance interpretation will then erroneously conclude that the the global system  is in a mixed state, not in a pure state as assumed in Equation~\eqref{Scho}). The error is assuming that the system is in a well-defined state that we (the humans) ignore it. This is simply  not true. D'Espagnat results shows that a conditional state cannot be a description of an open system with all the static and dynamic information that we can get from the open system. It does not mean that such states cannot still give some useful infomration of the system (as happens in Equation \eqref{result}).   

In addition, let us notice that  the conclusion of D’Espagnat applies to any open quantum system without distinguishing between Markovian or non-Markovian scenarios. However, indeed, there is no contradiction between the D’Espagnat conclusion and the attempt of the SSE of using pure states to describe Markovian open quantum systems for static and dynamic properties. Both are right. D'Espagnat discussion is a formal (fundamental) discussion about conditional states, while the discussion about Markovian scenarios is a practical discussion about simplifying approximation when extracting information of the system at large $\tau$. 

Finally, let us notice that the D'Espagnat conclusions does not apply to Bohmian mechanics becuase the Bohmian definition of a quantum system involves a wave function plus trajectories. The conditional state  $\tilde \psi_{Q^i(t)}(x,t)=\Psi(x,Q^i(t),t)$, together with the environment and system trajectory  $Q^i(t)$ and $X^i(t)$,  contains all the (static and dynamic) information of the of the open system in this $i$-th experiment. An ensemble over all  experiments prepared with the same global wavefunction $\Psi(x,q,t)$ requires an ensemble of different environment and system trajectories $Q^i(t)$ and $X^i(t)$ for $i=1,2,...,M$ with $M\to\infty$. But,  we repeat $\tilde \psi_{Q^i(t)}(x,t)=\Psi(x,Q^i(t),t)$ with  $Q^i(t)$ and $X^i(t)$ contains all the physical information on the subsystem, corresponding to  the $i$-th experiment.

\section{Orthodox and Bohmian reduced density matrices}
\label{mean_boh}

The Orthodox and Bohmian definitions of a quantum state are different. The first uses only a wave function, while the second uses the same wave function plus trajectories. It is well known that both reproduce the same ensemble values by construction. Here, we want to discuss how the orthodox density matrix (without trajectories) can be described by the Bohmian theory with trajectories. 

We consider a system plus environment defined by a Hilbert space $\mathcal{H}$ that can be decomposed as $\mathcal{H}=\mathcal{H}_x\otimes\mathcal{H}_q$ where $x$ is the collective positions of particles of the system while ${q}$ are the collective positions of the particles of the environment.The expectation value of any observable $O_x$ of the system can be computed as $\langle O_x\rangle = \langle \Psi|\hat{O}_x\otimes \hat{I}_{q}|\Psi\rangle$ with $\hat{I}_{q}$ the identity operator for the environment. We describe the typical orthodox procedure to define the orthodox reduced density matrix by tracing out all degrees of freedom of the environment:
\begin{equation}
    \rho({x},{x}\;',t)=\int d{q}\Psi^*({x}\;',{q},t)\Psi({x},{q},t)
    \label{density}
\end{equation}
From Equation~\eqref{density} the mean value of the observable $O_x$ can be computed as,
\begin{equation}
    \langle O_x \rangle=\int d{x}\left (O_x\rho(x,x',t)|_{{x}\;'={x}}\right)
    \label{mean}
\end{equation}
In this appendix, we want to describe Equation~\eqref{density} and Equation~\eqref{mean} in terms of the Bohmian conditional wavefunctions and trajectories described in the text. The conditonal wavefunction associated to the system during the $i$-th experiment conditioned on a particular value of the environment $Q^i(t)$ is defined as $\psi_{Q^i(t)}(x,t)=\Psi(x,Q^i(t),t)$, being $\Psi({x}, q,t)=\langle {x}, q|\Psi\rangle$ the position representation of the global state. We start from the general expression for the ensemble value in the position representation as,
\begin{eqnarray}
\langle O_x\rangle = \int dx \int dq \;\Psi^*(x,q,t) O_x\Psi(x,q,t)
\end{eqnarray}
Multiplying and dividing by $\Psi^*(x,q,t)$ we get,
\begin{eqnarray}
\langle O_x\rangle&=&\int dx \int dq \;\frac{|\Psi(x,q,t)|^2 O_x\Psi(x,q,t)}{\Psi^*(x,q,t)}\nonumber\\
&=&\frac{1}{M}\sum_{i=1}^M \int dx \int dq \frac{\delta[(x-X^i(t)]\delta[(q-Q^i(t)]O_x\Psi(x,q,t)}{\Psi^*(x,q,t)}\nonumber\\
&=&\frac{1}{M}\sum_{i=1}^M\int dx \frac{O_x\Psi(x,Q^i(t),t)}{\Psi^*(x,Q^i(t),t)}\delta(x-X^i(t))
\label{mean_2}
\end{eqnarray}
where we have used the quantum equilibrium condition $|\Psi(x,q,t)|^2=\frac{1}{M}\sum_{i=1}^M\delta[(x-X^i(t)]\delta[(q-Q^i(t)]$ with $M\to\infty$. Now, we multiply and divide by $\Psi^*(x,Q^i(t),t)$ to get,
\begin{eqnarray}
\langle O_x\rangle&=&\frac{1}{M}\sum_{i=1}^M\int dx \frac{O_x\Psi^*( {x},Q^i(t),t)\Psi(x,Q^i(t),t)}{|\Psi(x,Q^i(t),t)|^2}\delta(x-X^i(t))\nonumber\\
&=&\int dx\left[O_x\sum_{i=1}^M P_i\tilde{\psi}^{i*}(x',t)\tilde{\psi}^i(x,t)\right]_{x'=x=X^i(t)}
\label{MEAN_boh}
\end{eqnarray}
where $P_i=1/M$ can be interpreted as the probability associated to each $i=1,2,...,M$ experiment and we have defined: 
\begin{equation}
\tilde{\psi}^i(x,t)\equiv\frac{\Psi(x,Q^i(t),t)}{\Psi(X^i(t),Q^i(t),t)}. 
\end{equation}
Now, once we arrive at Equation~\eqref{MEAN_boh}, one can be tempted to define a type of Bohmian reduced density matrix in terms of the conditional wavefunctions for $i=1,2,...,M$ experiments as,
\begin{eqnarray}
    \rho_w(x\;',x,t)=\sum_{i=1}^M P_i\;\tilde{\psi}^{i*}(x\;',t)\tilde{\psi}^i(x,t)=\sum_{i=1}^M P_i\;\frac{\Psi^{*}(x\;',Q^i(t),t)}{\Psi^{*}(X^i(t),Q^i(t),t)}\frac{\Psi(x,Q^i(t),t)}{\Psi(X^i(t),Q^i(t),t)}
    \label{density_wrong}
\end{eqnarray}
where we have, arbitrarily eliminated the role of the trajectories. But strictly speaking Equation~\eqref{density} is not equal to Equation~\eqref{density_wrong}. If we include all $i=1,2,...,M$ experiments in the computation of \eqref{density_wrong}, there are trajectories $Q^i(t)$ and $Q^j(t)$ that at the particular time $t$ can be represented by the same conditional wavefunction $\tilde{\psi}^i(x,t)=\tilde{\psi}^j(x,t)$ if $Q^i(t)=Q^j(t)$. Such over-summation due to the repetition of the same trajectories is not present in \eqref{density}. 

To simplify the subsequent discussion, let us assume that $q$ is one degree of freedom in a 1D space. Let us cut such 1D space into small intervals of length $\Delta q$. Each interval is defined as $j\; \Delta q < q <(j+1)\; \Delta q $ and it is labelled by the index $j$. Then, we can define $G^j(t)$ as the number of positions $Q^i(t)$ that are inside the $j$-interval at time $t$ as: 
\begin{eqnarray}
    G^j(t)=\sum_{i=1}^M \int_{j\; \Delta q}^{(j+1)\; \Delta q} \delta[q-Q^i(t)] dq \label{norma}
\end{eqnarray}
With this definition, assuming that $\Delta q$ is so small that all $Q^i(t)$ inside the interval and all the corresponding Bohmian conditional wave functions $\Psi(x,Q^j(t),t)$, system positions $X^i(t)$ and probabilities  $P_i$ are almost equivalent, and given by $q^j$, $\Psi(x,q^j,t)$, $x^j$ and  $P_j$ respectively, we can change the sum over $i=1,...,M$ experiments into a sum over  $j=...,-1,0,1,...$ spatial intervals to rewrite Equation \ref{density_wrong} as:
\begin{eqnarray}
    \rho_w(x\;',x,t)=\sum_{j=-\infty}^{j=+\infty} G^j(t) P_j\;\frac{\Psi^{*}(x\;',q^j,t)}{\Psi^{*}(x^j,q^j,t)}\frac{\Psi(x,q^j,t)}{\Psi(x^j,q^j,t)} \approx
      \int dq^j N^j(t)\Psi^{*}(x\;',q^j,t)\Psi(x,q^j,t) 
    \label{density_wrong2}
\end{eqnarray}
where $N^j(t)=G^j(t) P_j\;/\left(\Psi^{*}(x^j,q^j,t)\Psi(x^j,q^j,t)\right)$. So, finally, a proper normalization of the Bohmian conditional states allows us to arrive to Equation~\eqref{density} from Equation~\eqref{density_wrong}. Such normalization is already discussed in the Equation~\eqref{B_cond_s} in the text. The moral of the mathematical developments of this appendix is that open systems are more naturally described in terms of density matrix than in terms of conditional states when using the orthodox theory, while the contrary happens when using the Bohmian theory. Because of the additional variables of the Bohmian theory, the conditional states are a natural Bohmian tool to describe open systems.    

\section{Equations of motion for single-electron conditional states in graphene} 
\label{graph}

As said in the text, graphene dynamics are given by the Dirac equation and not by the usual Schr\"{o}dinger one. The presence of the Dirac equation on the description of the dynamics of electrons in graphene is not due to any relativistic correction, but to the presence of a linear energy-momentum dispersion (in fact, the graphene Fermi velocity $v_f=10^6 \: m/s$ is faster than the electron velocity in typical parabolic band materials, but still some orders of magnitude slower than the speed of light). Thus, the conditional wavefunction associated to the electron is no longer a scalar, but a bispinor. In particular, the initial bispinor is defined (located outside of the active region) as: 
\begin{equation}
\label{bispinor}
\begin{pmatrix}
\psi_1(x,z,t) \\
\psi_2(x,z,t) 
\end{pmatrix}= \left(\begin{matrix}
1 \\ se^{i\theta_{\vec{k_c}}}
\end{matrix}\right)\Psi_g(x,z,t)
\end{equation}
where $\Psi_g(x,z,t)$ is a gaussian function with central momentum $\vec{k_c}=(k_{x,c},k_{z,c})$, $s=1$ ($s=-1$) if the electron is in the conduction band (valence band) and $\theta_{\vec{k_c}}=atan(k_{z,c}/k_{x,c})$. The wave packet can be considered as a Bohmian conditional wavefunction for the electron, a unique tool of Bohmian mechanics that allows to tackle the many-body and measurement problems in a computationally efficient way \cite{marian, colomes2017quantum}. The two components are solution of the mentioned Dirac equation:
\begin{eqnarray}
\label{dirac2d2}
i \hbar \frac{\partial }{\partial t}
\begin{pmatrix}
\psi_1\\ 
\psi_2 
\end{pmatrix}=
\begin{pmatrix}
V(x,z,t) & -i\hbar v_f \frac{\partial }{\partial x}-\hbar v_f \frac{\partial }{\partial z}\\
-i\hbar v_f \frac{\partial }{\partial x}+\hbar v_f \frac{\partial }{\partial z} & V(x,z,t)
\end{pmatrix} 
\begin{pmatrix}
\psi_1 \\
\psi_2 
\end{pmatrix}=
&-i\hbar v_f \left(\vec{\sigma} \cdot \vec{\nabla}+V\right)\left( \begin{array}{lcr}
\psi_1 \\
\psi_2 
      \end{array}
    \right)
\end{eqnarray}
where $v_f=10^6 \: m/s$ is the mentioned Fermi velocity and $V(x,z,t)$ is the electrostatic potential. $\vec{\sigma} $ are the Pauli matrices:
\begin{eqnarray}
\vec{\sigma}=(\sigma_x,\sigma_z)=
\left(
\begin{pmatrix}
0&1\\
1&0\\
\end{pmatrix}
,
\begin{pmatrix}
0&-i\\
i&0\\
\end{pmatrix}
\right)
\label{11}
\end{eqnarray}
Usually, in the literature, one finds $\sigma_z$ as $\sigma_y$, however, since we defined the graphene plane as the $XZ$ one, the notation here is different. Then, we can obtain a continuity equation for the Dirac equation and then we can easily identify the Bohmian velocities of electrons as \cite{pladevall2019applied}
\begin{equation}
\vec{v}(\vec{r},t)=\dfrac{ v_f\psi(\vec{r},t)^{\dagger}\vec{\sigma}\psi(\vec{r},t)}{|\psi(\vec{r},t)|^2}
\label{bvel}
\end{equation}
By time integrating \eqref{bvel} we can obtain the quantum Bohmian trajectories. The initial positions of the trajectories must be distributed according to the modulus square of the initial wavefunction, i.e., satisfying the quantum equilibrium hypothesis and thus certifying the same empirical results as the orthodox theory \cite{pladevall2019applied}. All this formalism was introduced in the BITLLES simulator in order to correctly model graphene and other linear band structure materials.

\bibliography{bibliography}

\begin{thebibliography}{74}
\bibitem [{foo()}]{footnote1}%
  \BibitemOpen
  \href@noop {} {}\bibinfo {note} {We highlight the (nonlocal) dependence of
  the Bohmian velocities of the particles of the environment on the particles
  of the system, and vice-versa. This shows just the entanglement between
  environment and system at the level of the Bohmian trajectories.}\BibitemShut
  {Stop}%
\end{thebibliography}
 
\end{document}